\documentclass[preprint,letterpaper,twocolumn,11pt]{article}
\usepackage[english]{babel}
\usepackage[latin1]{inputenc}
\usepackage{blindtext}
\usepackage{authblk} %para agregar +info del ator
\usepackage{amsmath}
\usepackage{amsfonts}
\usepackage{amssymb}
\usepackage{makeidx}
\usepackage{enumerate} % for lists
\usepackage{cite} % para citar las referencias
 \usepackage{url}
\usepackage{booktabs} % para el diseño de la tabla
\usepackage{array, multicol}
\usepackage{caption}
\usepackage{wrapfig}
\usepackage{subcaption}
\usepackage{graphicx}
\usepackage{etoolbox} % espacio entre las columnas
\usepackage{setspace} % para sangría y espacio entre párrafos
\usepackage[colorlinks=true, citecolor=blue, linkcolor=blue, urlcolor=blue]{hyperref}
\usepackage[top=1.0in,bottom=1.0in,right=0.65in,left=0.65in]{geometry} %color de las citas

\usepackage[runin]{abstract}
\abslabeldelim{:\quad} % add a colon to the abstract title

\title{\textbf{Single-pixel imaging with Fourier filtering: application to vision through scattering media}}

\author[1,*]{Y. Jauregui-S\'anchez} %https://orcid.org/0000-0002-3720-685X
\author[2]{P. Clemente}
\author[1]{J. Lancis}
\author[1]{E. Tajahuerce} %https://orcid.org/0000-0003-1655-4393

\affil[1]{GROC$\cdot$UJI, Institute of New Imaging Technologies (INIT), Universitat Jaume I, 12071, Castell\'o, Spain}
\affil[2]{Servei Central d'Instrumentaci\'o Cient\'ifica (SCIC), Universitat Jaume I, 12071, Castell\'o, Spain}
\affil[*]{Corresponding author: jauregui@uji.es}

\date{}

\begin{document}

\twocolumn[
\vspace*{-0.5cm}
\begin{@twocolumnfalse}
\maketitle
\begin{center}\rule{0.9\textwidth}{0.1mm} \end{center}
\begin{abstract}
We present a novel approach for imaging through scattering media that combines the principles of Fourier spatial filtering and single-pixel imaging. We compare the performance of our single-pixel imaging setup with that of a conventional system. Firstly, we show that a single-pixel camera does not reduce the frequency content of the object, even when a small pinhole is used as a low-pass filter. Second, we show that the introduction of Fourier gating improves the contrast of imaging through scattering media in both optical systems. We conclude that single-pixel imaging fits better than conventional imaging on imaging through scattering media by Fourier gating. \\

\begin{center}\rule{0.9\textwidth}{0.1mm} \end{center}
\vspace*{0.75cm}
\end{abstract}
\end{@twocolumnfalse}
]
 
 \singlespacing %espacio entre líneas 
%\doublespacing 

Imaging through scattering media has been a longstanding issue in many applications in engineering and biomedical imaging. Different approaches to tackle the problem can be classified as a function of the regime of light used to reconstruct the image \cite{Ntziachristos10, Dunsby03}. Diffuse optical techniques, such as diffuse optical tomography (DOT), relay on totally diffused photons. In this macroscopic regime, when light has travel several transport mean free paths (TMFP), images are obtained by solving inverse problems based on diffuse models \cite{Zhang03}. Recent techniques based on the control of light wavefronts with spatial light modulators (SLM) have allowed to develop imaging methods in an intermediate regime of light where photons are only slightly dispersed. These techniques allows to work in a mesoscopic regime, reaching an intermediate depth range in addition to a good resolution \cite{Gigan17,Mosk12}. Finally, some techniques for optical imaging through scattering media extract information of hidden objects from unscattered ballistic light. These ballistic techniques provide the best resolution although, of course, paying the price of a low penetration depth. Since, in general, the scattered component of the light is stronger than the ballistic component, these techniques aim to reduce the scattered one in order to improve the signal to noise ratio of the final image. For instance, time-gating techniques select ballistic photons of light pulses by taking into account that they arrive earlier to the detector \cite{Wang91}. Other procedures to discriminate ballistic against scattered light are based on the use of some form of spatial filtering, by considering that most photons are scattered to higher spatial frequencies \cite{Dunsby03}. Fourier filtering is a practical and effective technique that, combined with ultrafast time-gating imaging, has allowed to reduce the scattered light by 10 orders of magnitude \cite{Alfano95}. Recently high-contrast images through scattering media have been obtained employing both Fourier filtering and structured illumination \cite{Berrocal16}. However, the use of a spatial filter to reduce the scattered light limits the achievable spatial resolution of the final image since high frequencies are blocked. In fact, there is a tradeoff between image resolution and contrast inherent to this technique. In this work we show how this tradeoff can be overcome by combining Fourier filtering and single-pixel imaging methods.

The first single pixel imaging (SPI) techniques were described already in 1970 \cite{Decker70}. However, efficient cameras based on SPI were developed only recently \cite{Duarte08} by using fast programmable spatial light modulators (SLM). SPI techniques are characterized by using structured illumination and bucket detection. The object is sequentially sampled with a set of microstructured light patterns codified onto an SLM, for instance a digital micromirror device (DMD). The light transmitted (or reflected) by the object is recorded by a single photosensor such as a photodiode or a photomultiplier tube (PMT). The image is reconstructed in the computer from the electric signal properly digitized by a data acquisition system (DAQ). To this end, different mathematical algorithms can be used, such as a simple linear superposition, a change of basis transformation, or a correlation operation. Single-pixel imaging (SPI) has proved to be an effective technique for imaging through scattering media \cite{Taja14, Duran15, Chen15}.

In this work, we describe an optical system working by transmission that combines SPI and the principles of Fourier spatial filtering to recover the image of an object hidden behind a turbid medium. We compare the performance of our optical setup with that of a conventional imaging system based on a CMOS camera. Our results show that introducing Fourier spatial filtering improves the contrast of the images in both cases. However, the resolution of the conventional imaging system decreases with the amount of spatial frequency gating, while in the case of SPI the resolution loss is negligible. We conclude that SPI fits better than conventional imaging in vision through scattering media by Fourier filtering.

%\section{Single-pixel imaging with Fourier filtering}

\begin{figure}[h!]
\captionsetup{singlelinecheck = false, labelfont=bf, font=small}
  \begin{subfigure}{9.0cm}
    \caption{}
     \centering
    \includegraphics[width=8.6cm]{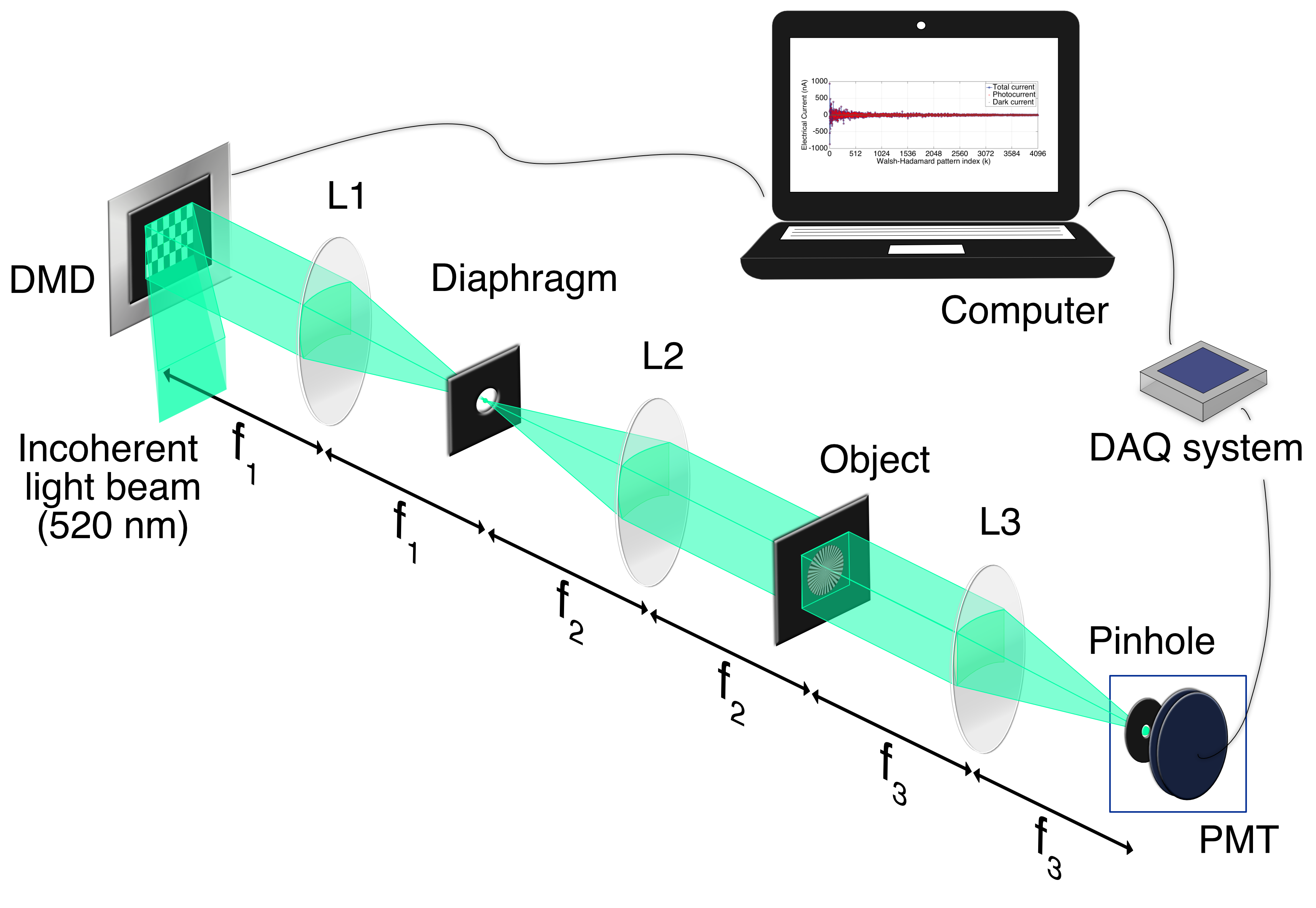}
  \end{subfigure}
      \hfill
    \begin{subfigure}{9.0cm}
    \caption{}
     \centering
    \includegraphics[width=8.6cm]{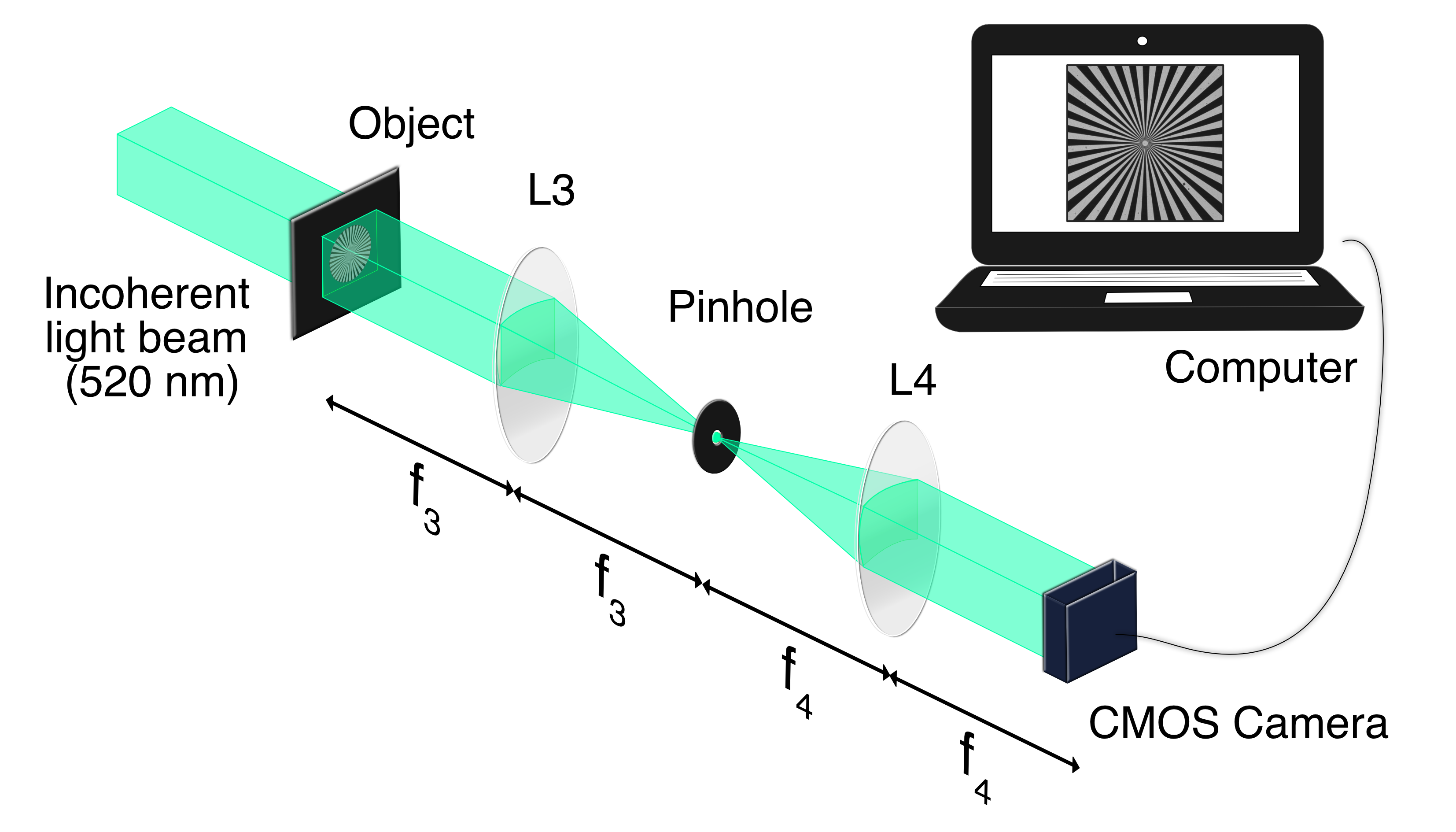}
  \end{subfigure}
     \caption{ Schematic diagram of (a) the single-pixel camera and (b) the conventional imaging setup.}
  \label{fig:1}
\end{figure}

The experimental configurations to study the performance of the single-pixel camera with Fourier filtering is shown in Fig. \ref{fig:1}(a). The microstructured light patterns are generated with a DMD illuminated with a monochromatic collimated light beam. An optical system in a 4-f configuration, constituted by lenses L1 and L2, projects the patterns onto the object. A circular diaphragm is used to filter unwanted diffracted orders produced by the periodic micromirror arrangement on the DMD. Light is collected by lens L3 and focused onto a PMT. A pinhole with variable diameter located at the back Fourier plane of L3 performs the spatial frequency gating. With the aim of comparing, we use a second experimental setup constituted by a conventional 4-f imaging system, formed by lenses L3 and L4, as is shown in Fig. \ref{fig:1}(b). Images are recorded by a conventional camera. Again, a pinhole located at the Fourier plane filters the spatial frequencies of the input object.

%----------------------------------------------------- IMAGE 1 -----------------------------------
%----------------------------------------------------- END-----------------------------------------

In our experiments, the DMD (DLP Discovery 4100, Texas Instrument) is a chipset array constituted by 1920 x 1080 micromirrors with a pixel pitch of 10.8 $\mu m$ and a display resolution of 1080p. The patterns codified on the DMD are 2D functions of the orthonormal Walsh-Hadamard (WH) basis \cite{Pratt69} with 64 x 64 pixels. The light beam is generated with an incoherent while-light source. An interference filter (P10-515S 93819, Corion) with a bandwidth of 20 nm centered at 520 nm is used to avoid spectral artifacts. The focal distances of L1, L2 and L3 are the same, $f_1 = f_2 = f_3 =$ 100 mm, while that of L4 is $f_4$ = 75 mm. The detector in Fig. \ref{fig:1}(a) is a PMT (PMM01, Thorlabs Inc.) and the sensor in Fig. \ref{fig:1}(b) is a CMOS camera (UI-1540SE-M-GL, IDS) with 1280 x 1024 pixels with a pitch of 5.2 $\mu m$. The magnification factor of the 4-f optical system in Fig. \ref{fig:1}(b) is 0.75.

\begin{figure}[t]
\captionsetup{singlelinecheck = false, labelfont=bf, font=small}
     \centering
      \includegraphics[width=8.5cm]{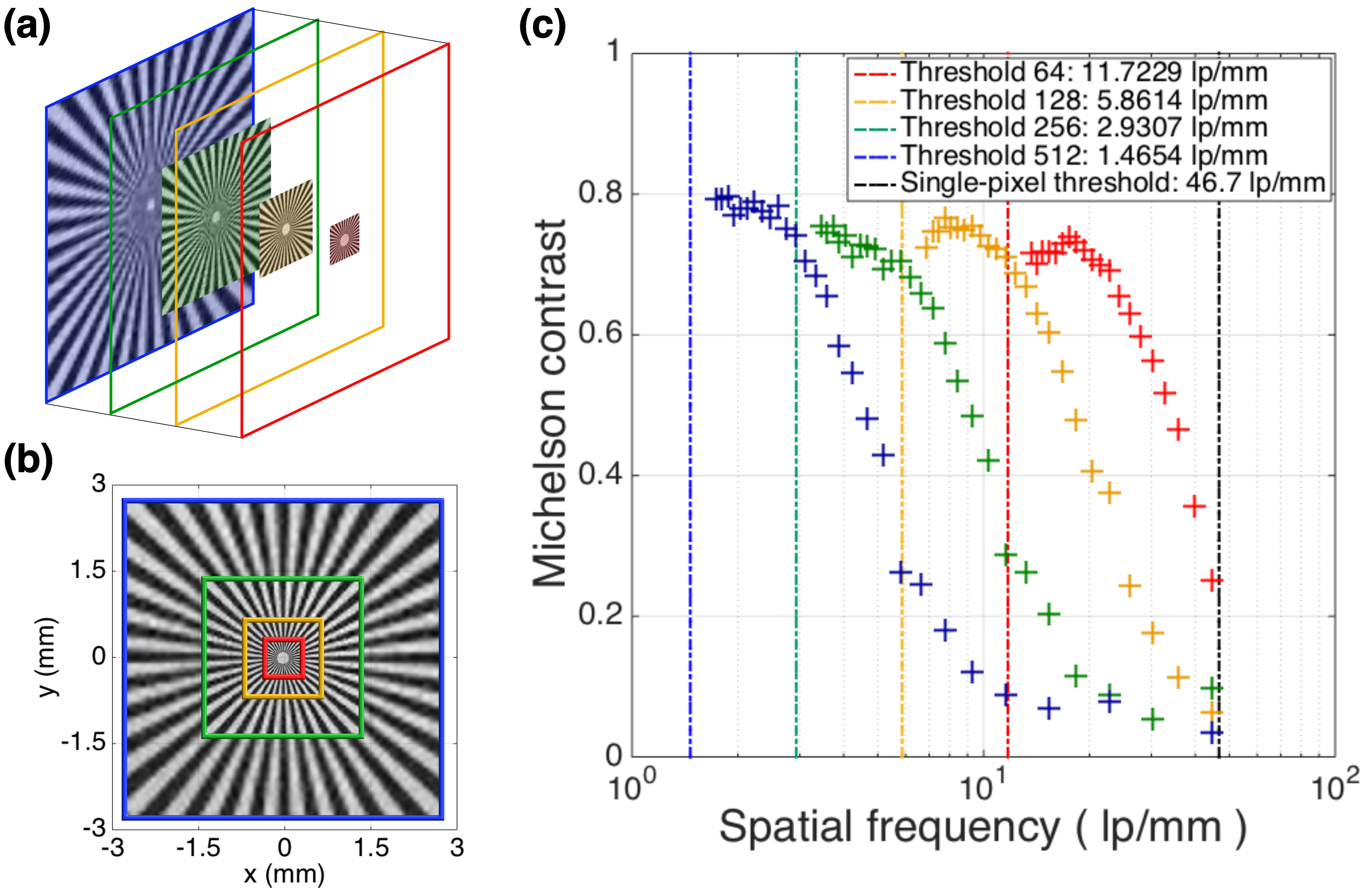}
     \caption{Experimental results obtained with the optical system in Fig.\ref{fig:1}(a) with no Fourier filtering: (a) Images of the sector star target recorded with different resolutions; (b) Final image obtained by combining the elemental images in (a); (c) Michelson contrast as a function of the spatial frequency for the elemental images in (a).}
  \label{fig:2}
\end{figure}

The object in both experiments is a sector star target R1L1S2P \cite{Thorlabs} that contains 36 radial sector pairs (opaque and transparent). To operate with the single pixel camera obtaining the maximum resolution, a wide field of view (FOV), and the shortest possible measurement time, we adopt the following strategy, which resembles that used in \cite{Padgett17}. We project different sets of microstructured patterns, all of them with a low number of pixels but with different sizes, adapting the spatial resolution of the sampling patterns to that of the object. Our approach is related with methods that use signal processing techniques to obtain high-resolution images from multiple low-resolution samples \cite{Park03}. In particular, we project 4 sets of Hadamard patterns each one consisting on 64 x 64 pixels codifying the full collection of 4096 WH functions for these resolution. Each pixel of the pattern is codified with a number of micromirrors that varies from 8 x 8, for the set with lower spatial resolution, to 1 x 1, for the patterns with higher spatial resolution. By projecting these patterns onto the object, using the optical system in Fig. \ref{fig:1}(a), and reconstructing the images with SPI techniques, we obtain four elemental images with 512 x 512, 256 x 256, 128 x 128, and 64 x 64 pixels as is shown in Fig. \ref{fig:2}(a). The final image of the object is generated by combining the previous images into a single one, overlapping the images with higher resolution over those with lower resolution, as is shown in Fig. \ref{fig:2}(b). Note that, by using this strategy, we recover a single image with 512 x 512 pixels but with a reduced number of projected patterns compared with those required in conventional SPI techniques. The number of projected patterns and, accordingly, the measurement time, is reduced by a factor of 16, equal to the ration of 512 x 512 to 4 x 64 x 64. However, the image has the maximum spatial resolution at the center, where the object has smaller details, and just lower spatial resolution at the borders, where the object shows low frequencies. In Fig. \ref{fig:2}(c) we plot the Michelson contrast, or visibility, of the 4 elemental images in Fig. \ref{fig:2}(a) as a function of the spatial frequency. For each elemental image, the threshold of the spatial resolution, which is marked with a vertical color line for each curve in Fig. \ref{fig:2}(c), corresponds to the pixel size of the corresponding WH pattern codified on the DMD.

%----------------------------------------------------- IMAGE 2 -----------------------------------
%----------------------------------------------------- END-----------------------------------------

\begin{figure}[h!]
\captionsetup{singlelinecheck = false, labelfont=bf, font=small}
 \begin{subfigure}{9.0cm}
   \caption{}
    \centering
   \includegraphics[width=8.8cm]{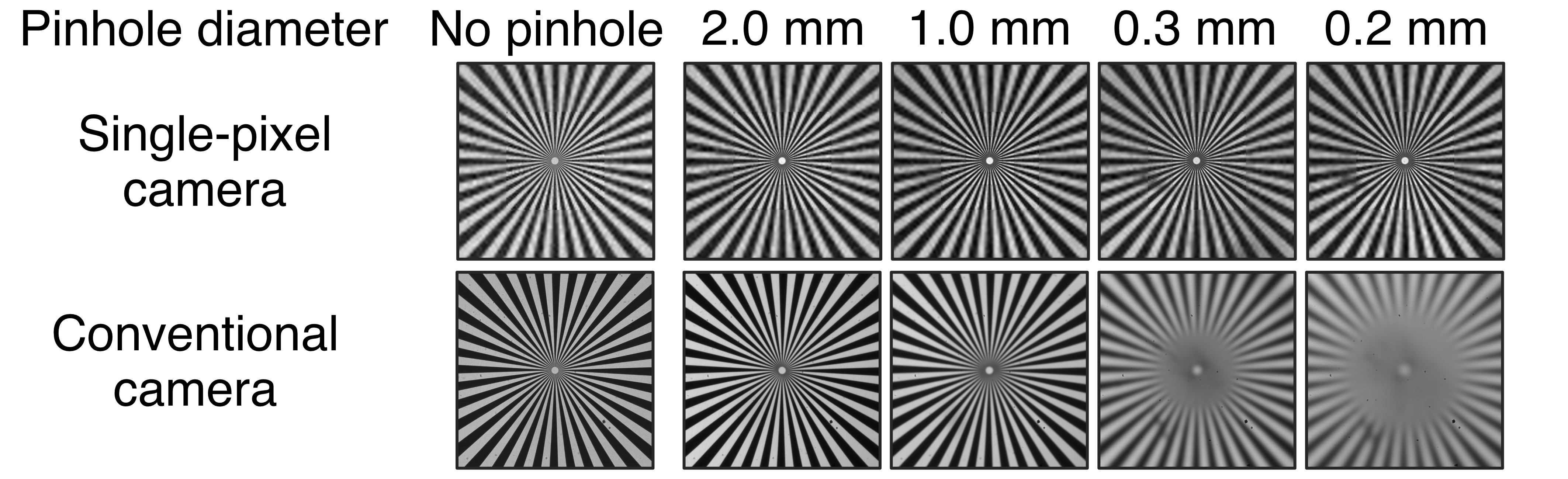}
 \end{subfigure}
        \hfill
 \begin{subfigure}{9.0cm}
  \caption{}
   \centering
  \includegraphics[width=7.3cm]{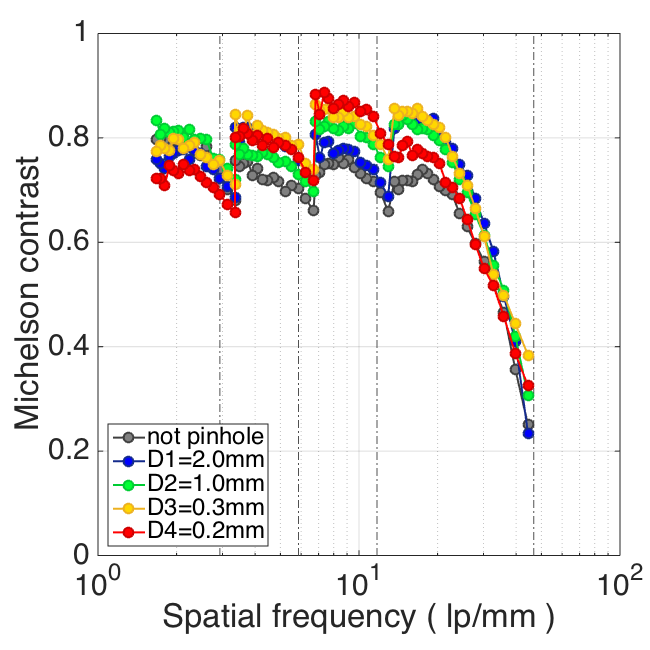}
 \end{subfigure}
        \hfill
 \begin{subfigure}{9.0cm}  
   \caption{}
    \centering
   \includegraphics[width=7.3cm]{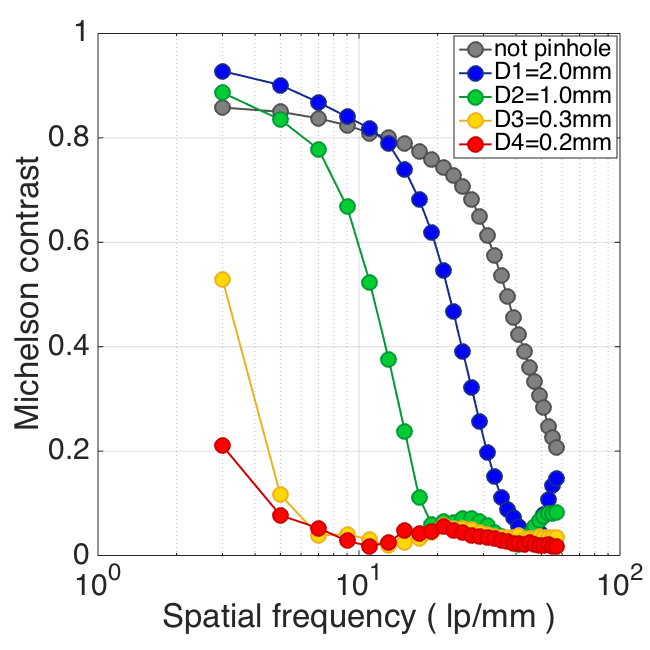}
 \end{subfigure}
     \caption{Experimental results obtained with the optical systems in Fig. \ref{fig:1} and Fourier filtering techniques. (a) Images recorded by both cameras and different pinhole diameter size on Fourier plane. Michelson contrast versus spatial frequency for (b) the single-pixel camera and (c) the conventional imaging system.}
  \label{fig:3}
\end{figure}

To analyze the influence of Fourier filtering on SPI we use a pinhole with variable diameter on the back Fourier plane of lens L3, as shown in Fig. \ref{fig:1}(a). Besides, we locate a pinhole with the same characteristics at the Fourier plane of the conventional imaging system in Fig. \ref{fig:1}(b). The experimental results are shown in Fig. \ref{fig:3}. Images in the first row of Fig. \ref{fig:3}(a) were obtained with the single-pixel camera, while those in the second row were recorded with the conventional one. Images on each column correspond to different Fourier filtering conditions: no pinhole for the first column and pinholes with a diameter of 2.0 mm, 1.0 mm, 0.3 mm, and 0.2 mm for the subsequent columns, respectively. We conclude that the loss in resolution is negligible when performing Fourier spatial filtering in the SPI approach, even for small pinhole diameters. On the contrary, the loss is clearly noticeable in the conventional imaging system as the pinhole diameter decreases. These experimental results were corroborated by plotting the Michelson contrast as a function of the spatial frequency. Figures \ref{fig:3}(b) and \ref{fig:3}(c) show the results for the single-pixel camera and the conventional camera, respectively. Note that, by reducing the pinhole diameter, the contrast remains approximately constant for the case of SPI while it decreases significantly for the conventional imaging system. This behavior of the SPI system is due to the characteristics of the measurement process. In SPI, the image is reconstructed by measuring the projection of the object onto a set of test functions, in our case functions of the WH basis. The key point is that this information is already present in the zero order of the Fourier transform of the product of the sampling patterns with the object. The object is then obtained by a simple linear superposition of the functions of the basis with the measured coefficient even if we filter all frequencies but the zero order at the Fourier plane. The only effect of the filtering process is a reduction of the light power efficiency. 

%----------------------------------------------------- IMAGE 3 -----------------------------------
%----------------------------------------------------- END-----------------------------------------

%\section{Application to Imaging through Scattering Media}

A relevant application of Fourier filtering is on imaging through scattering media. It has been used to reject scattered light, thus increasing the proportion of ballistic light and, consequently, the image contrast \cite{Alfano95,Berrocal16}. However, this technique may also degrade the image quality, because the spatial-frequency gating operation will reduce also the spatial-frequency bandwidth of the final image. In view of the results obtained in the previous experiment, we propose to combine SPI techniques with Fourier spatial filtering to increase the quality of images of an object obtained through a turbid layer in a transmission configuration. To study the validity of the approach, we locate a scattering layer against the object in the SPI setup in Fig. \ref{fig:1}(a). The influence of the location of the scattering medium in SPI has been widely discussed in \cite{Taja14, Duran15, Rodriguez14}. It has been proved that a turbid media located after the object (between the object and the sensor) has no influence in the quality of the image. However, if the turbid media is located before the object (between the SLM and the object) the light patterns projected onto the object are degraded by scattering and the quality of the final image decreases. Therefore, in our experiment we choose the worse condition for SPI and locate the scattering layer before the object in the optical system in Fig. \ref{fig:1}(a). Again, we use a pinhole with variable diameter at the back Fourier plane of lens L3. For comparison, we repeat the experiment with the conventional system in Fig. \ref{fig:1}(b). In this case, the scattering layer is located after de object, as in \cite{Berrocal16}. As turbid medium we use a layer of epoxy resin and (TiO$_2$) rutile powder with a thickness of 3.27 mm. The turbid layer was made following the recipe of epoxy resin phantoms described in \cite{USAJauregui}. We mixed 0.725 (g/l) of TiO$_2$ rutile powder (Titanium (IV) oxide, rutile powder, $<5$ $\mu$m particle size, Sigma Aldrich\textregistered) \cite{TiO2} with a 1:2 ratio of hardener (component B) to epoxy resin (component A).

\begin{figure}[h!]
\captionsetup{singlelinecheck = false, labelfont=bf, font=small}
 \begin{subfigure}{9.0cm}
   \caption{}
    \centering
   \includegraphics[width=8.8cm]{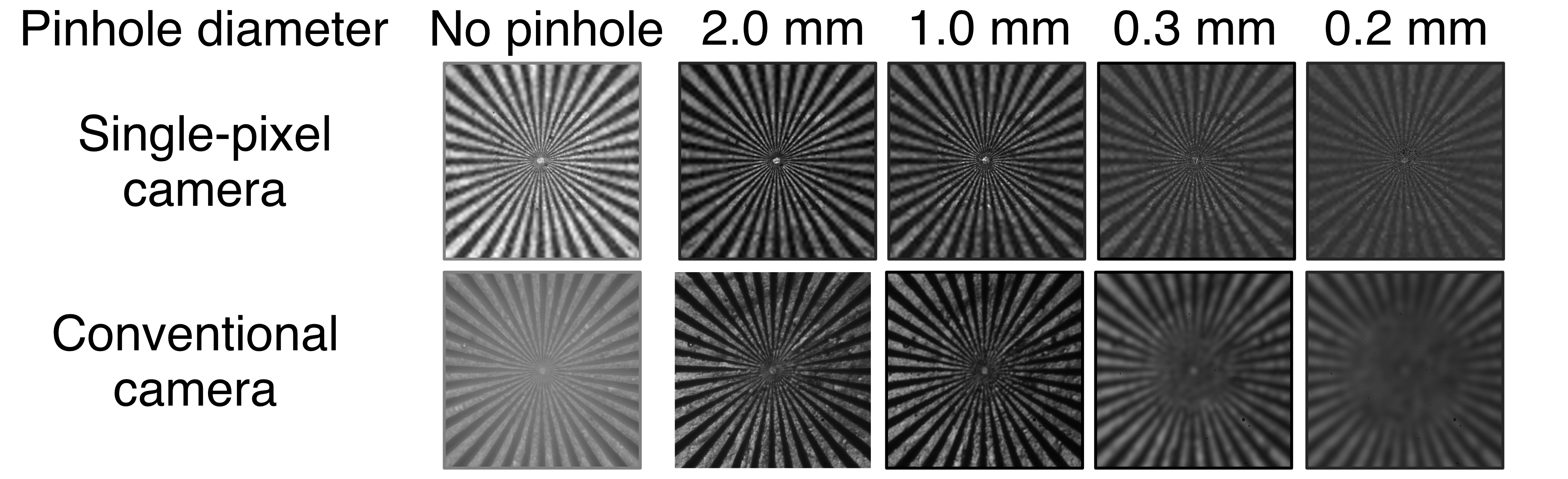}
 \end{subfigure}
        \hfill
 \begin{subfigure}{9.0cm}
  \caption{}
   \centering
  \includegraphics[width=7.3cm]{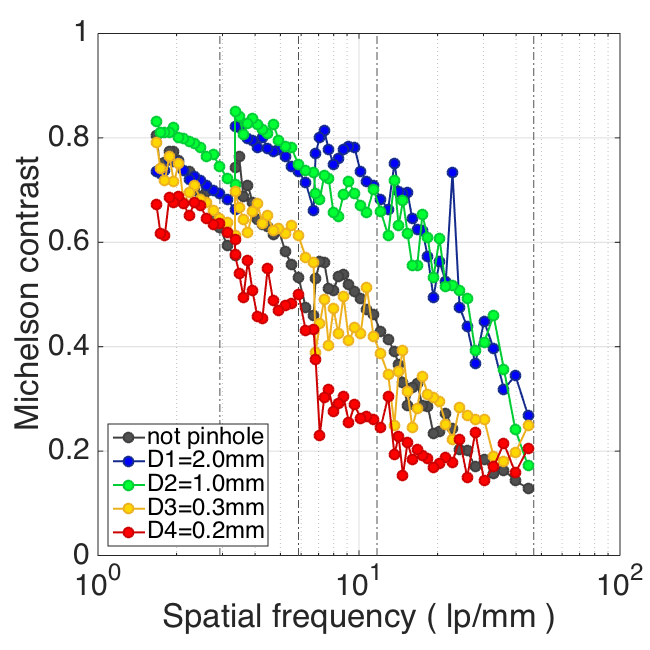}
 \end{subfigure}
        \hfill
 \begin{subfigure}{9.0cm}  
   \caption{}
    \centering
   \includegraphics[width=7.3cm]{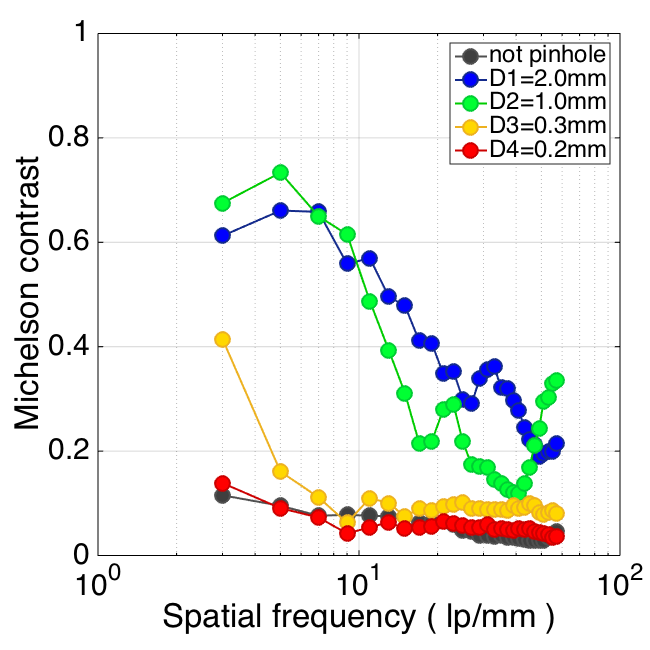}
 \end{subfigure}
     \caption{Experimental results with turbid medium. (a) Images obtained by both cameras in different Fourier filtering conditions. The spatial frequency curves versus the Michelson contrast for (b) the single-pixel camera, and (c) conventional optical imaging system.}
  \label{fig:5}
\end{figure}

The experimental results are shown in Fig. \ref{fig:5}. The different images are obtained in the same conditions as those in Fig. \ref{fig:3} but now with the scattering layer located against the object in the optical systems shown in Fig. \ref{fig:1}. In both cameras, the introduction of Fourier filtering improves the contrast of the images, as can be seen comparing the result in the second column with those in the first one in Fig. \ref{fig:5}(a). This effect was corroborated by evaluating the Michelson contrast as a function of the spatial frequency of the images provided by both cameras. The curves are shown in Figures \ref{fig:5}(b) and \ref{fig:5}(c) for the case of the SPI system and the conventional one, respectively. Note that, in both cases, the resolution system worsens as the pinhole diameter decreases further. In the case of the conventional imaging system, the main reason is the low-pass filtering effect of the spatial frequencies, as was shown in Fig. \ref{fig:3}. In the case of SPI, the image resolution deteriorates because the SNR of the photocurrent provided by the PMT decreases for small diameters of the pinhole. However, the contrast of the images provided by the SPI system is better than that of the conventional one for all Fourier filtering conditions, including the no filtering case.

% Figure 4 was eliminated

%----------------------------------------------------- IMAGE 5 -----------------------------------
%----------------------------------------------------- END-----------------------------------------

%\section{Conclusions}

\vspace{1.5cm}

In conclusion, we have developed an imaging system combining SPI techniques with Fourier spatial filtering. We have compared the performance of our optical setup with that of a conventional imaging system based on Fourier filtering and using a CMOS camera as detector. We have shown that, without scattering media, the single-pixel camera does not reduce the high frequency content of the object, even when a small pinhole is used as a low-pass filter. Moreover, when the scattering media is introduced, the Fourier spatial filtering technique improves the contrast of the images for both the single-pixel camera and the conventional one. We conclude that SPI fits better than conventional imaging in vision through turbid media by Fourier filtering. We would like to note that this effect is present in many SPI configurations using a photosensor with a reduced size. Therefore, it may contribute to improve the image quality in other applications of SPI in scattering media.

\section*{Funding Information}

The authors acknowledge the financial support from MINECO (FIS2016-75618-R and FIS2015-72872-EXP), Generalitat Valenciana (PROMETEO/2016/079), and Universitat Jaume I (P1-1B2015-35). Yessenia Jauregui-S\'anchez acknowledges the Santiago Grisol\'ia support from Generalitat Valenciana (GRISOLIA/2015/037).

% ------------------------------------------------- COMPLETE REFERENCES ------------------------------------------------

\end{document}